\documentclass[runningheads]{llncs}
\usepackage[T1]{fontenc}
\usepackage{amsmath,amssymb}
\usepackage{booktabs,tabularx,array,multirow}
\usepackage{xcolor}
\usepackage{url}
\usepackage{graphicx}
\usepackage{enumitem}
\usepackage{tikz}
\usetikzlibrary{positioning,arrows.meta,decorations.pathreplacing}
\setlist{nosep}
\newcolumntype{Y}{>{\raggedright\arraybackslash}X}
\newcommand{\method}{\textsc{EC3}}

\AtBeginDocument{%
  \ifdefined\pdfpagewidth
    \pdfpagewidth=210mm
    \pdfpageheight=297mm
  \fi
  \pagestyle{empty}%
}

\begin{document}

\title{Evidence-Carrying Contracts for Safe Claim Composition in Data Marketplace Workflows}
\titlerunning{Evidence-Carrying Contracts for Data Marketplace Workflows}

\author{Duc-Trung Bui\inst{1} \and Phuong-Chi Le\inst{1} \and Quyet-Thang Huynh\inst{2}}
\authorrunning{D.-T. Bui et al.}
\institute{ICT Department, FPT University, Hanoi, Vietnam\\
\email{TrungBD9@fe.edu.vn}\\
\email{chilp2@fe.edu.vn} \and
School of ICT, Hanoi University of Science and Technology, Ha Noi, Vietnam\\
\email{thang.huynhquyet@hust.edu.vn}}

\maketitle

\begin{abstract}
Data marketplace workflows cross organizational boundaries, where payment,
policy presence, or API delivery may be promoted into transaction claims that
the evidence does not support. We introduce evidence-carrying claim contracts:
a lightweight assume--guarantee semantics declaring assumptions, guarantees,
blocking obligations, and evidence dependencies. A conflict-sensitive
four-state domain distinguishes supported, refuted, unresolved, and conflicting
requirements. Safe composition bounds the transaction claim by the meet of
local bounds and blocks full promotion while a critical obligation remains
open. We establish local soundness, compositional boundedness, revocation
safety, termination, and aggregation-order invariance for finite
acyclic workflows. An executable conformance harness covers all
$4^6=4{,}096$ configurations, 18 mutations, six basic revocations, 12
alternative-support scenarios, an eight-transition evidence trace, 2,880
aggregation schedules, 96 requirement-level and six interface checks. An
independent operational evaluator agrees with the declarative oracle;
mutation-sensitive campaigns detect all eight seeded defects, though no single
campaign detects all. It issues no unsupported full claim, preserves a full
claim under valid alternative support, and blocks it when none remains. Four
diagnostic negative controls, not competitive baselines, issue 63--1,023
unsupported full claims. These results establish executable conformance to
declared contracts, not specification completeness, evidence authenticity, or
legal validity.

\end{abstract}

\section{Introduction}

Data marketplaces and data spaces are not ordinary file stores. They coordinate providers, consumers, operators, reviewers, identity and access services, payment services, and post-use controls. A product may be discoverable without being licensable for every purpose; a payment may succeed while a use-right assumption remains unresolved; an API may enforce access while provenance is disputed; and a derived product may require rights not conveyed by the source license. Consequently, a transaction-level statement such as ``this exchange is admissible'' cannot be justified by a single local event.

This paper studies a narrow software-assurance question: \emph{how can claims produced by individual stages be composed without becoming stronger than their supporting evidence?} The question precedes legal certification, security testing, data valuation, and business optimization. It concerns the semantic entitlement of a workflow to emit a claim.

Existing contract-based design and assume--guarantee reasoning provide a foundation for compositional verification \cite{benveniste2018contracts,clarke1989compositional}. Provenance models explain how outputs depend on sources \cite{green2007provenance,prov2013}. Multi-valued and bilattice logics distinguish incomplete from inconsistent information \cite{belnap1977,ginsberg1988}. However, marketplace workflow descriptions commonly leave these concerns separated: a policy is present, an access check passes, or a payment completes, and the workflow label is promoted without an explicit cross-stage evidence contract.

The novelty is a claim-entitlement semantics jointly governing cross-stage
claim bounds, conflicts, blocking obligations, alternative support, and
revocation. We address three research questions:
\begin{description}[leftmargin=1.0cm,style=nextline]
\item[RQ1] How can stage assumptions, guarantees, obligations, and evidence dependencies be represented as evidence-carrying contracts?
\item[RQ2] Under what conditions can local claims be composed without producing a stronger unsupported transaction claim?
\item[RQ3] Does an executable evaluator prevent unsafe full-claim promotion under missing, refuted, conflicting, expired, and revoked evidence?
\end{description}

\begin{sloppypar}
The contributions are: (i) an evidence-carrying claim contract
$\mathcal H_i=\langle \mathsf{Asm}_i,\mathsf{Gua}_i,\mathsf{Obl}_i,
\mathsf{Dep}_i,\Theta_i,\mathsf{Pol}_i\rangle$ (Eq.~\ref{eq:hi}) with a
four-state evidence interpretation; (ii) safe composition, four propositions,
one lemma, and explicit claim boundaries; and (iii) an executable artifact with
an evaluator independent of its oracle and mutation-sensitive campaigns. The
method checks declared contracts and evidence; it does not establish evidence
truth, legal completeness, or that a marketplace is secure, fair, or compliant.
\end{sloppypar}

\section{Background and Related Work}

\subsection{Data marketplaces, data spaces, and governance}

Data marketplaces mediate discovery, contracting, delivery, and monetization of data products \cite{stahl2017marketplaces,spiekermann2019data,ramadan2024trading}, and recent systems address cross-domain exchange, access policies, privacy-preserving computation, and compliant distributed AI transactions \cite{mavrogiorgou2022crossdomain,more2022compute,gheisari2025compliant}. Data-space architectures emphasize sovereignty, usage control, interoperability, trust, and provenance \cite{otto2022dataspaces,curry2022dataspaces,idram2022,moller2024ecosystems,scherenberg2024sovereignty}, clarifying components and roles but not providing a rule for bounding a transaction-level claim by heterogeneous stage evidence. W3C PROV provides an interoperable vocabulary for entities, activities, agents, and derivations \cite{prov2013}; we use provenance as a dependency carrier inside claim contracts rather than proposing a new provenance standard.

\begin{sloppypar}
Reference architectures make the operational context concrete: human-centric architectures connect consent, catalogues, payment, purpose limitation, processing, and revocation \cite{scheider2023reference}; the DSSC Blueprint and Gaia-X Trust Framework organize interoperable governance and trust services \cite{dssc2024blueprint,gaiax2022trust}; and the Dataspace Protocol specifies catalogue publication, agreement negotiation, and transfer processes \cite{dsp2025}. These govern important parts of an exchange, whereas our question is whether the evidence behind their local outcomes semantically entitles a workflow to emit a stronger end-to-end claim.
\end{sloppypar}

\subsection{Contracts, composition, and conflicting evidence}

Assume--guarantee reasoning decomposes a system proof into component assumptions and guarantees \cite{clarke1989compositional}. Contract-based design generalizes this view to system engineering and interface refinement \cite{benveniste2018contracts}. Runtime verification monitors temporal properties of executions \cite{leucker2009runtime}, while temporal logic specifies safety and liveness over traces \cite{pnueli1977}. Our contribution is not a new general contract calculus; it specializes contract composition to evidence-governed claims, so that contracts carry evidence dependencies and blocking obligations and composition is ordered by claim strength. Four-valued logics distinguish true, false, both, and neither \cite{belnap1977}; bilattices separate truth and knowledge orders \cite{ginsberg1988}. Paraconsistent treatment is valuable when different sources support and refute the same admission-critical requirement. We use a four-state domain operationally: unresolved and conflicting requirements cannot be silently treated as passing, and conflict does not explode into arbitrary conclusions.

\subsection{Positioning}

Table~\ref{tab:position} summarizes the gap at the level of declared capabilities: a check means the capability is a primary, explicit object of the cited approach, whereas ``partial'' means related information exists but is not used as a cross-stage claim-admission rule. This paper contributes a small assurance layer connecting these lines; it does not replace policy languages, access-control systems, provenance stores, or legal review.

\begin{table}[tb]
\caption{Relationship to adjacent approaches.}\label{tab:position}
\centering\scriptsize
\setlength{\tabcolsep}{3pt}
\begin{tabularx}{\textwidth}{@{}p{2.45cm}Yccccc@{}}
\toprule
Approach & Primary concern & Bound & Conflict & Oblig. & Revoke & Trace \\
\midrule
Marketplace/data-space architectures \cite{scheider2023reference,moller2024ecosystems} & Roles, services, governance, and exchange flow & -- & -- & Partial & Partial & Partial \\
Policy and privacy-preserving markets \cite{more2022compute} & Pointwise permission and protected computation & -- & -- & Partial & -- & Partial \\
Gaia-X/DSSC trust frameworks \cite{gaiax2022trust,dssc2024blueprint} & Federated trust, compliance, and interoperability & -- & Partial & Partial & Partial & Partial \\
Dataspace Protocol \cite{dsp2025} & Catalogue, agreement, and transfer interoperability & -- & -- & Partial & Partial & Partial \\
Provenance and contract reasoning \cite{prov2013,benveniste2018contracts} & Dependencies and component-level composition & Partial & -- & Partial & -- & $\checkmark$ \\
\method{} & Evidence-governed cross-stage claim composition & $\checkmark$ & $\checkmark$ & $\checkmark$ & $\checkmark$ & $\checkmark$ \\
\bottomrule
\end{tabularx}
\end{table}

\section{Motivating Workflow and Requirements}

\subsection{Six-stage workflow}

We construct an illustrative data-marketplace workflow by synthesizing recurring functions in public sources: cross-domain marketplace workflows \cite{mavrogiorgou2022crossdomain}, human-centric data-ecosystem architectures \cite{scheider2023reference}, the IDS Reference Architecture Model \cite{idram2022}, the DSSC Blueprint \cite{dssc2024blueprint}, and the Dataspace Protocol \cite{dsp2025}. The six stages are a conformance fixture, not a transcription of one proprietary requirements report and not empirical evidence of deployment effectiveness; they retain only the functions needed to study cross-stage claim composition.

\emph{Registration} carries provider identity, product description, and asserted
supply right; \emph{Review} carries provenance, sensitivity, and quality
evidence; \emph{Listing} carries approval, catalog publication, and bounded
valuation claims; \emph{License} carries consumer purpose, use conditions,
duration, and derived-use rights; \emph{Access} carries payment status,
authorization, API delivery, and access logs; and \emph{Post-use} carries purpose
monitoring, derivation, redistribution, expiry, revocation, and dispute.

% ---------------------------------------------------------------- Figure 1
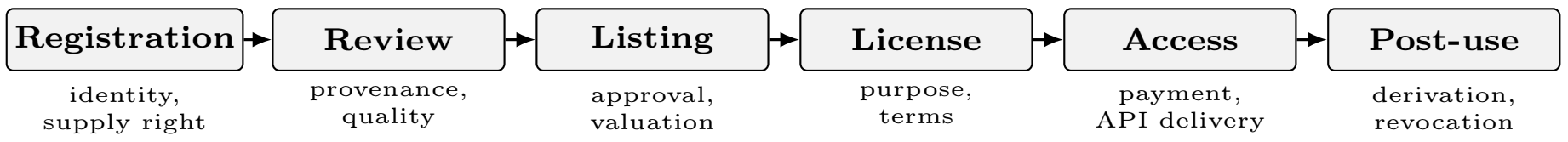
\begin{figure}[tb]
\centering
\begin{tikzpicture}[
  x=2.05cm,
  stage/.style={draw, rounded corners=1.5pt, fill=black!5,
                minimum height=4.8mm, minimum width=18mm, align=center,
                inner xsep=0.6mm, inner ysep=0.9mm,
                font=\scriptsize\bfseries},
  ev/.style={font=\tiny, align=center, text width=18mm, inner sep=0pt},
  ar/.style={-{Latex[length=1.5mm,width=1.2mm]}, semithick}
]
\node[stage] (s1) at (0,0) {Registration};
\node[stage] (s2) at (1,0) {Review};
\node[stage] (s3) at (2,0) {Listing};
\node[stage] (s4) at (3,0) {License};
\node[stage] (s5) at (4,0) {Access};
\node[stage] (s6) at (5,0) {Post-use};
\foreach \a/\b in {s1/s2, s2/s3, s3/s4, s4/s5, s5/s6}
  \draw[ar] (\a) -- (\b);
\node[ev, below=1.1mm of s1] {identity,\\supply right};
\node[ev, below=1.1mm of s2] {provenance,\\quality};
\node[ev, below=1.1mm of s3] {approval,\\valuation};
\node[ev, below=1.1mm of s4] {purpose,\\terms};
\node[ev, below=1.1mm of s5] {payment,\\API delivery};
\node[ev, below=1.1mm of s6] {derivation,\\revocation};
\end{tikzpicture}
\caption{The six-stage workflow used as the executable conformance fixture.
Each arrow is a contract interface $\mathcal H_i\triangleright\mathcal H_{i+1}$,
not evidence that the next stage is entitled to emit a full claim; the text
under each stage names its admission-critical evidence.}\label{fig:workflow}
\end{figure}

\subsection{Failure patterns}

\begin{sloppypar}
The workflow exposes five recurring promotion errors: $\mathsf{PaymentSuccess}$ $\not\Rightarrow$ $\mathsf{UseEntitled}$; $\mathsf{PolicyDeclared}$ $\not\Rightarrow$ $\mathsf{PolicyEnforced}$; $\mathsf{Catalogued}$ $\not\Rightarrow$ $\mathsf{SupplyRightVerified}$; $\mathsf{UseRight}(d)$ $\not\Rightarrow$ $\mathsf{ShareRight}(\mathsf{derive}(d))$; and local passage does not imply global admissibility while an upstream obligation remains open. These are semantic non-implications, not legal conclusions: a jurisdiction or contract may add rules, but a stronger right cannot be inferred from the listed local event alone.
\end{sloppypar}

\section{Evidence-Carrying Claim Contracts}

\subsection{Evidence states}

For each admission-critical requirement $q$ of stage $i$, supporting and refuting evidence are aggregated separately over the active evidence profile $\eta_t$. Let $L_q^+(\eta_t)$ and $L_q^-(\eta_t)$ be their support levels: $L_q^+$ by a \emph{bottleneck} rule --- the weakest usable (active, in-scope, valid) member of a declared support set, maximized over alternative sets, so any one sufficient alternative suffices --- and $L_q^-$ by the strongest usable refuting item. Let $\tau_{i,q}^+$, $\tau_{i,q}^-$ be the declared thresholds. Define
\begin{equation}
b_{q,t}^+=\mathbf{1}[L_q^+(\eta_t)\geq\tau_{i,q}^+],\qquad
b_{q,t}^-=\mathbf{1}[L_q^-(\eta_t)\geq\tau_{i,q}^-].
\label{eq:threshold}
\end{equation}
The evidence state is
\begin{equation}
\mathsf{state}_t(q)=
\begin{cases}
\mathbf{U} & (b_{q,t}^+,b_{q,t}^-)=(0,0),\\
\mathbf{T} & (b_{q,t}^+,b_{q,t}^-)=(1,0),\\
\mathbf{F} & (b_{q,t}^+,b_{q,t}^-)=(0,1),\\
\mathbf{C} & (b_{q,t}^+,b_{q,t}^-)=(1,1),
\end{cases}
\label{eq:state}
\end{equation}
where $\mathbf{U}$, $\mathbf{T}$, $\mathbf{F}$, and $\mathbf{C}$ denote unresolved, supported/true, refuted/false, and conflicting. A requirement passes only in state $\mathbf{T}$ and only if scope and freshness conditions also hold, i.e.\ its supporting evidence is active, in scope, and within its validity window (see the bottleneck rule above).

\subsection{Contracts, claims, and obligations}

A stage contract is
\begin{equation}
\mathcal H_i=\langle \mathsf{Asm}_i,\mathsf{Gua}_i,\mathsf{Obl}_i,\mathsf{Dep}_i,\Theta_i,\mathsf{Pol}_i\rangle,
\label{eq:hi}
\end{equation}
where $\mathsf{Asm}_i$ contains assumptions, $\mathsf{Gua}_i$ guarantees, $\mathsf{Obl}_i$ unresolved obligations, $\mathsf{Dep}_i$ evidence dependencies, $\Theta_i$ a typing context, and $\mathsf{Pol}_i$ a claim policy of permitted and prohibited claims. Stages here carry a single flat claim family, so $\Theta_i$ is instantiated at the trivial one-type context ($\Theta_i:=\top$) and $\mathsf{Pol}_i$ coincides with the claim-bound map $\beta_i$ below; the typed setting is left to future work. For each declared claim family $f$, claim strength is a finite meet-semilattice
\begin{equation}
\mathbf C_f=(\mathcal{C}_f,\preceq_C,\wedge_C,\bot_f),
\label{eq:lattice}
\end{equation}
where $\mathcal C_f$ is the carrier set, $c_1\preceq_C c_2$ means that $c_1$ is no stronger than $c_2$, every finite nonempty set has a greatest lower bound under $\wedge_C$, and $\bot_f$ is the least element, denoting that no positive transaction claim is emitted. This structure is local to a declared claim family; we do not assume one universal ordering across unrelated legal or business claims.

A contract is locally admissible at time $t$ iff
\begin{equation}
\mathsf{LocalAdm}(\mathcal H_i,t)\iff
\bigwedge_{q\in \mathsf{Asm}_i}\mathsf{pass}_{\mathcal H_i}(q,\eta_t)
\land \mathsf{Obl}_i^{\mathrm{block}}(t)=\varnothing.
\label{eq:localadm}
\end{equation}

The contract emits one of four decisions:
\begin{itemize}[leftmargin=*]
\item \textsc{Admissible}: all critical requirements are supported;
\item \textsc{Conditional}: none is refuted/conflicting but at least one is unresolved;
\item \textsc{Prohibited}: at least one is refuted and none is conflicting; or
\item \textsc{Abstain}: at least one critical requirement is conflicting.
\end{itemize}
Conflict has priority because it requires source resolution before a stronger conclusion is safe. For comparing output strength in $\mathcal C_f$, define the claim-bound map $\beta_i$:
\begin{equation}
\begin{aligned}
\beta_i(\textsc{Admissible})&=c_i^{\max}, &
\beta_i(\textsc{Conditional})&=c_i^{\mathrm{cond}}\preceq_C c_i^{\max},\\
\beta_i(\textsc{Prohibited})&=\bot_f, &
\beta_i(\textsc{Abstain})&=\bot_f.
\end{aligned}
\label{eq:beta}
\end{equation}
This mapping orders positive claim strength; it does not equate the operational meanings of prohibition and abstention.

\subsection{Safe composition}

Adjacent contracts are interface-compatible when
\begin{equation}
\mathcal H_i\triangleright \mathcal H_{i+1}
\iff \mathsf{Gua}_i\models \mathsf{Asm}_{i+1}\land \mathsf{Obl}_i^{\mathrm{block}}=\varnothing.
\label{eq:interface}
\end{equation}
For a workflow $\mathcal{W}=\mathcal H_1\otimes\cdots\otimes\mathcal H_n$, the transaction claim must satisfy
\begin{equation}
c_{\mathcal{W}}\preceq_C\bigwedge_{i=1}^{n}c_i,
\label{eq:bound}
\end{equation}
and its outstanding obligations are
\begin{equation}
\mathsf{Obl}_{\mathcal{W}}=\left(\bigcup_{i=1}^{n}\mathsf{Obl}_i\right)\setminus\mathsf{Discharged}(\mathcal{W}).
\label{eq:oblw}
\end{equation}
The meet in Eq.~\ref{eq:bound} is a claim bound; it does not assert that every claim family forms one universal linear order.

\subsection{Revocation}\label{sec:revocation}

Each evidence item has a validity interval and may be revoked. Let $\mathcal E$ be the evidence universe and $\mathsf{Act}_t\subseteq\mathcal E$ the active, scope-valid evidence at time $t$. For a claim $c$, let $\mathsf{SuppSets}(c)\subseteq 2^{\mathcal E}$ be its declared family of alternative sufficient dependency sets: $c$ is evidence-supported only if some $X\in\mathsf{SuppSets}(c)$ satisfies $X\subseteq\mathsf{Act}_t$ and its associated requirements pass. Sets in $\mathsf{SuppSets}(c)$ may overlap; an independent alternative support set need not contain the revoked item. If $\mathsf{DepClaims}(e)$ is the transitive set of claims having at least one declared dependency set containing $e$, the operational bounded-response requirement is
\begin{equation}
\mathbf{G}\!\left(\mathsf{revoked}(e)\rightarrow
\mathbf{F}_{\leq\Delta_{\mathrm{rev}}}\!\bigwedge_{c\in\mathsf{DepClaims}(e)}
\left(\neg\mathsf{Supported}_{t'}(c)\rightarrow\neg\mathsf{permitted}_{t'}(c)\right)\right),
\label{eq:revoke}
\end{equation}
where $t'$ is the bounded future time reached by $\mathbf{F}_{\leq\Delta_{\mathrm{rev}}}$, distinct from the outer $\mathbf{G}$-time at which $\mathsf{revoked}(e)$ holds. $\mathsf{Supported}_{t'}(c)$ holds iff an active $X\in\mathsf{SuppSets}(c)$ satisfies all declared support conditions. The bound is $\Delta_{\mathrm{rev}}=0$ for the synchronous evaluator and may be a declared service-level bound in deployment. Here $\mathsf{permitted}(c)$ is recomputed from all sets in $\mathsf{SuppSets}(c)$, so revoking one item does not invalidate a claim still supported by another active sufficient set. A deployment must state whether this bounded-response specification is read in metric temporal logic or an equivalent timed transition system.

\emph{Relation to provenance-derived support sets.} The declared family $\mathsf{SuppSets}(c)$ need not be minimal or exhaustive. A provenance algebra would derive $\mathsf{prov}(c)=\bigoplus_j\bigotimes_{e\in S_j}x_e$ and the inclusion-minimal witnesses $\mathsf{MinSupp}(c)$; the two coincide exactly under $\mathsf{Closed}(c)\iff\mathsf{SuppSets}(c)=\mathsf{MinSupp}(c)$. Proposition~3 and Lemma~1 hold for the declared family regardless, but $\mathsf{MinSupp}(c)$ may replace it only where $\mathsf{Closed}(c)$ is separately established, a conservativity obligation this evaluator does not discharge.

\section{Assurance Properties and Evaluator}

Intuitively, each stage is a checkpoint, and the end-to-end claim cannot exceed
the weakest checkpoint. Missing evidence lowers the claim, conflicting evidence
forces abstention, and revocation triggers re-evaluation. The properties below
formalize these safeguards; the evaluator checks their implementation.

\subsection{Properties}

\textbf{Proposition 1 (Local contract soundness).}
If $\mathcal H_i$ emits a full claim, every critical assumption of that claim is in state $\mathbf{T}$, satisfies its scope/freshness constraints, and has no blocking obligation.

\emph{Justification.} This follows directly from the definition of $\mathsf{LocalAdm}$ and the full-decision branch. The proposition is relative to the declared critical set; it does not prove that the contract author captured every real-world requirement.

\textbf{Proposition 2 (Compositional boundedness).}
For a compatible workflow $\mathcal{W}=\mathcal H_1\otimes\cdots\otimes\mathcal H_n$, the emitted workflow claim satisfies Eq.~\ref{eq:bound}.

\emph{Justification.} Composition applies the meet operator to local claim bounds after checking every interface. By the defining property of meet, the result is no greater than any operand. Any blocking obligation diverts evaluation from the full branch.

\textbf{Lemma 1 (Monotonic safety under evidence degradation).}
Fix the contracts, dependency families, thresholds, and all requirement states except $q$. If $q$ changes from $\mathbf{T}$ to $\mathbf{U}$, $\mathbf{F}$, or $\mathbf{C}$ without introducing new active evidence, then the reevaluated positive claim $c'$ cannot be stronger than the previous claim $c$: $c'\preceq_C c$.

\emph{Justification.} Local admission predicates are conjunctions of passing critical requirements. Replacing one passing state by $\mathbf{U}$ yields at most $c_i^{\mathrm{cond}}\preceq_C c_i^{\max}$; replacing it by $\mathbf{F}$ or $\mathbf{C}$ maps the output to $\bot_f$. Unchanged stages retain their bounds, and meet is no greater than any operand, so composition cannot increase the claim. This is a safety property, not monotonicity in the bilattice knowledge order.

\textbf{Proposition 3 (Dependency-set revocation non-persistence).}
After revoking $e$, a claim $c$ may remain full only if there exists an alternative $X\in\mathsf{SuppSets}(c)$ such that $e\notin X$, $X\subseteq\mathsf{Act}_{t+1}$, and all corresponding requirements still pass. If every sufficient $X\in\mathsf{SuppSets}(c)$ contains $e$, or otherwise becomes invalid after transitive dependency propagation, then $c$ is not emitted as full at the next synchronous evaluation.

\emph{Justification.} Revocation removes $e$ from $\mathsf{Act}_{t+1}$, invalidating every sufficient set containing it, and support is recomputed existentially over the remaining members of $\mathsf{SuppSets}(c)$. An independent valid set preserves entitlement, whereas the absence of any valid set makes some critical requirement non-passing, and Lemma~1 blocks persistence of a stronger claim. An omitted dependency is a specification-completeness defect, not a violation the evaluator detects.

\textbf{Proposition 4 (Termination and aggregation-order invariance).}
Fix a finite acyclic workflow $\mathcal{W}=\mathcal H_1\otimes\cdots\otimes\mathcal H_n$, an evidence profile, and the workflow-order interface checks. Deterministic local evaluation terminates, and for fixed local bounds $c_1,\ldots,c_n$ every parenthesization and permutation of the schedule computing $\bigwedge_{i=1}^{n}c_i$ yields the same workflow claim bound. This concerns only the aggregation schedule; it does not permit reordering the contracts or changing the workflow edges.

\emph{Justification.} A topological evaluation of a finite acyclic workflow visits each stage once over finite input and terminates. Once local bounds are fixed, associativity makes parenthesization irrelevant, commutativity makes the schedule irrelevant, and idempotence prevents a repeated bound from changing the result; none of these laws changes the ordered relation $\mathsf{Gua}_i\models\mathsf{Asm}_{i+1}$.

\emph{Three assurance boundaries.} Propositions 1--4 and Lemma~1 establish only
\emph{contract-relative soundness}: for the declared requirements, dependencies,
thresholds, and evidence, no emitted claim exceeds its bounds. Enumeration cannot
establish \emph{specification completeness}, which remains a modeling question.
Nor does a passing guard prove \emph{evidence authenticity}; that requires
signatures, attestations, or source audits.

\subsection{Executable evaluator}

Given ordered contracts $\mathcal H_1,\ldots,\mathcal H_n$ and an active evidence profile $\eta_t$, the evaluator (i) aggregates each critical requirement's declared alternative support sets and computes its positive and negative threshold indicators; (ii) assigns $\mathbf{U}$, $\mathbf{T}$, $\mathbf{F}$, or $\mathbf{C}$; (iii) evaluates local claim status and blocking obligations; (iv) checks each interface $\mathsf{Gua}_i\models\mathsf{Asm}_{i+1}$; and (v) abstains with conflict-resolution obligations if any conflict exists, else prohibits the full claim if any refutation exists, else returns a conditional claim if any requirement is unresolved, else returns the meet-bounded full claim. It outputs a decision, obligations, and an evidence trace. For the fixed workflow used here, runtime is linear in the number of stages and requirements; we do not generalize this bound to recursive or dynamically generated contract languages.

\section{Evaluation}

\subsection{Design}

The evaluation is a conformance study of the declared semantics. Each of the six stages receives one of four states, creating $4^6=4{,}096$ configurations. The oracle is a declarative reading of the decision table on stage states: conflict $\rightarrow$ abstain; otherwise refutation $\rightarrow$ prohibited; otherwise unresolved $\rightarrow$ conditional; otherwise full. The evaluator is a separate operational pipeline that never consults it: given claim contracts and an evidence profile, it aggregates each requirement's declared support sets, derives $\mathbf{U}$, $\mathbf{T}$, $\mathbf{F}$, or $\mathbf{C}$ from the thresholds of Eq.~\ref{eq:state}, evaluates $\mathsf{LocalAdm}$, checks every interface, and folds the local bounds over $\mathbf{C}_f$. This triangulates the declarative specification, its independent operational implementation, and mutation-sensitive campaigns; it checks conformance, not predictive accuracy.

We additionally test:
\begin{itemize}[leftmargin=*]
\item 18 single mutations changing each stage of an all-supported workflow to $\mathbf{U}$, $\mathbf{F}$, or $\mathbf{C}$; six revocation sequences, each revoking one previously supported stage; and 12 dependency-set scenarios covering independent and overlapping alternatives, expiry, scope invalidity, conflict, and transitive propagation, plus an eight-transition evidence trace;
\item 2,880 aggregation-schedule checks: all $6!$ orders in which the six fixed local decisions are folded for each of four representative fixed workflows, with contract identities, edges, and interface checks held fixed. The fold operator is separately checked for idempotence, commutativity, and associativity, and an order-dependent variant is run as a negative control;
\item 96 requirement-level checks placing both requirements of each stage in all 16 ordered state pairs, which exercises the within-stage branch priority that stage-level enumeration cannot observe;
\item six interface checks, one compatible and five with an unsatisfied $\mathsf{Gua}_i\models\mathsf{Asm}_{i+1}$; and eight seeded implementation defects, each perturbing only the evaluator, to measure whether these campaigns detect a defective evaluator at all.
\end{itemize}

We include four \emph{diagnostic negative controls}. They are intentionally
incomplete counterexample policies, not competitive baselines or
representatives of the state of the art. Each isolates one promotion error that the contract semantics is designed to block.
\textbf{D1 Payment-success} emits a full transaction claim whenever the
access/payment stage is supported, omitting all other stage requirements.
\textbf{D2 Unknown-as-pass} emits a full claim when no stage is refuted or
conflicting, treating unresolved evidence as sufficient. \textbf{D3
Policy-presence} emits a full claim when no stage is unresolved, collapsing
refuted and conflicting evidence into mere policy presence. \textbf{D4
Conflict-collapse} emits a full claim whenever no stage is refuted, treating
unresolved and conflicting states as non-blocking.

The primary measure is the number of \emph{unsafe full claims}: a method emits \textsc{Admissible} while the oracle does not. We also report full-claim precision, overall exact agreement, revocation persistence failures, aggregation-order violations, and seeded-defect detection. No significance tests are used because the experiment exhaustively enumerates the finite declared state space.

\subsection{Results}

The oracle assigns one configuration to \textsc{Admissible}, 63 to \textsc{Conditional}, 665 to \textsc{Prohibited}, and 3,367 to \textsc{Abstain}. Table~\ref{tab:results} reports the diagnostic comparison. Every shortcut policy exposes unsupported full-claim promotions under its deliberately omitted guard. D1 is the most permissive, with 1,023 unsafe full claims (24.98\% of the state space). D2 promotes all 63 configurations containing only supported/unresolved stages. D3 and D4 each emit 728 unsafe full claims: each excludes exactly one of the four per-stage states (D3: $\mathbf{U}$; D4: $\mathbf{F}$), so both admit $3^6=729$ configurations, coinciding numerically though the admitted sets differ. These counts quantify the consequence of each isolated omission within the declared fixture; they do not establish empirical superiority over external systems. The proposed evaluator emits one full claim, the all-supported configuration, and no unsafe full claim.

\begin{table}[tb]
\caption{Exhaustive results on 4,096 evidence configurations. ``Precision'' is precision of the full-claim decision, not predictive precision on sampled field data.}\label{tab:results}
\centering\small
\begin{tabular}{@{}lrrrr@{}}
\toprule
Policy & Admissible issued & Unsafe full & Unsafe/all & Precision \\
\midrule
D1 Payment-success & 1,024 & 1,023 & 24.98\% & 0.10\% \\
D2 Unknown-as-pass & 64 & 63 & 1.54\% & 1.56\% \\
D3 Policy-presence & 729 & 728 & 17.77\% & 0.14\% \\
D4 Conflict-collapse & 729 & 728 & 17.77\% & 0.14\% \\
\method{} proposed & 1 & 0 & 0.00\% & 100.00\% \\
\bottomrule
\end{tabular}
\end{table}

All 18 single mutations produce the prescribed downgrade under \method{}. In the revocation campaign, revoking an indispensable item removes support without producing refuting evidence, so the affected stage becomes unresolved ($\mathbf U$) rather than refuted ($\mathbf F$). \method{} has zero full-claim persistence failures, whereas D2 and D4 fail all six sequences because they read $\mathbf U$ as passing, and D1 fails five. D3 fails none, but only incidentally: it blocks on unresolved evidence, exactly the state revocation produces, while still promoting refuted and conflicting evidence (728 unsafe full claims in Table~\ref{tab:results}), so surviving this campaign does not indicate dependency tracking. Across 2,880 aggregation-schedule checks \method{} has zero violation; these permute only the fold schedule over fixed local decisions, not stages or interfaces.

Exhaustive agreement is informative only if a defective evaluator would be caught, so we seed eight implementation defects into the operational pipeline and leave the oracle untouched. Table~\ref{tab:defects} shows that every defect is detected and that no single campaign detects all of them. Perturbing $\tau^+$ or $\tau^-$, replacing the bottleneck aggregation of Eq.~\ref{eq:threshold} with a best-member rule, and folding with an order-dependent operator are caught by the exhaustive enumeration. Demoting each stage's second requirement from critical and permuting the branch priority are invisible to it, because stage-level enumeration never places two distinct non-supported states inside one stage; these are caught only at requirement level, and dropping the interface check only by the interface cases. The three campaigns are therefore non-redundant, and the exhaustive agreement of the unmodified evaluator is a discriminating result rather than a restatement of the oracle. The fold operator satisfies idempotence, commutativity, and associativity, while the order-dependent control produces 360 schedule violations.

\begin{table}[tb]
\caption{Seeded implementation defects: disagreements with the oracle per campaign.}\label{tab:defects}
\centering\small
\begin{tabular}{@{}lrrr@{}}
\toprule
Seeded defect & Exh.\ (4,096) & Req.\ (96) & Itf.\ (6) \\
\midrule
$\tau^+$ lowered below the bottleneck & 728 & 48 & 0 \\
$\tau^-$ made unreachable & 4,032 & 72 & 0 \\
Best member replaces bottleneck & 728 & 24 & 0 \\
Second requirement not critical & 0 & 36 & 0 \\
Interface check dropped & 0 & 0 & 5 \\
Refutation ordered before conflict & 0 & 12 & 0 \\
Unresolved ordered before refutation & 0 & 24 & 0 \\
Order-dependent fold operator & 2,303 & 0 & 0 \\
\bottomrule
\end{tabular}
\end{table}

The artifact separates the declarative oracle from the operational evaluator, and includes all result tables and summary JSON. Exhaustive enumeration took about 0.5 seconds on the evaluation machine (four runs ranged 0.37--0.58s); this is descriptive, not a scalability claim.

\subsection{Alternative support and evidence trace}\label{sec:trace}

We exercise Proposition~3 with independent sufficient sets
\[
X_1=\{e_{registry},e_{lineage}\},\quad X_2=\{e_{audit},e_{custody}\},
\]
and overlapping sets $X_3=\{e_a,e_b\}$ and $X_4=\{e_b,e_c\}$. Across 12 scenarios covering revocation, expiry, scope invalidity, conflict, and a transitive downstream claim, the evaluator agrees with the dependency-set oracle in all 12, with zero alternative-support preservation failures and zero indispensable-support persistence failures.

% Figure 2: claim strength along the evidence trace
\begin{figure}[tb]
\centering
\begin{tikzpicture}[
  x=1.36cm, y=0.62cm,
  lvl/.style={font=\tiny, anchor=east, inner sep=1pt},
  tick/.style={font=\tiny, align=center, text width=13mm, inner sep=0pt},
  pt/.style={circle, draw, fill=black!70, inner sep=0pt, minimum size=1.5mm},
  grid/.style={draw=black!18, semithick},
  ev/.style={font=\tiny, inner sep=0.6pt, fill=white, text=black!65}
]
% decision levels, strongest at the top
\node[lvl] at (-0.18,3) {\textsc{Admissible}};
\node[lvl] at (-0.18,2) {\textsc{Conditional}};
\node[lvl] at (-0.18,1) {\textsc{Abstain}};
\node[lvl] at (-0.18,0) {\textsc{Prohibited}};
\foreach \y in {0,1,2,3} \draw[grid] (0,\y) -- (7,\y);
\draw[decorate, decoration={brace, amplitude=1.2mm}, black!45]
  (7.12,1.22) -- (7.12,-0.22) node[midway, right=0.9mm, font=\tiny, black!60]
  {$\bot_f$};
% claim path t0..t7
\draw[-{Latex[length=1.4mm,width=1.1mm]}, black!80, thick]
  (0,2) -- (1,2) -- (2,2) -- (3,3) -- (4,3) -- (5,2) -- (6,1) -- (7,0);
\foreach \x/\y in {0/2, 1/2, 2/2, 3/3, 4/3, 5/2, 6/1, 7/0}
  \node[pt] at (\x,\y) {};
% support-set annotations on the path
\node[ev, above=0.7mm] at (1,2) {$X_1,X_2$ active};
\node[ev, above=0.7mm] at (4,3) {$X_1$ revoked, $X_2$ holds};
\node[ev, below=0.7mm] at (5,2) {$X_2$ revoked};
% time axis
\foreach \x/\lab in {0/{$t_0$\\register}, 1/{$t_1$\\lineage,\\audit},
                     2/{$t_2$\\review,\\listing}, 3/{$t_3$\\payment,\\delivery},
                     4/{$t_4$\\lineage\\revoked}, 5/{$t_5$\\audit\\revoked},
                     6/{$t_6$\\sources\\conflict}, 7/{$t_7$\\license\\expired}}
  \node[tick, anchor=north] at (\x,-0.42) {\lab};
\end{tikzpicture}
\caption{Claim strength along the eight-transition evidence trace of
Sec.~\ref{sec:trace}: the claim never exceeds its active support, reaching
\textsc{Admissible} only at $t_3$, surviving revocation of $X_1$ at $t_4$ via
the independent set $X_2$, and falling back once the last sufficient set is
lost ($t_5$), sources conflict ($t_6$), or the license expires ($t_7$).
\textsc{Prohibited} and \textsc{Abstain} both map to $\bot_f$ but carry
different obligations.}\label{fig:trace}
\end{figure}
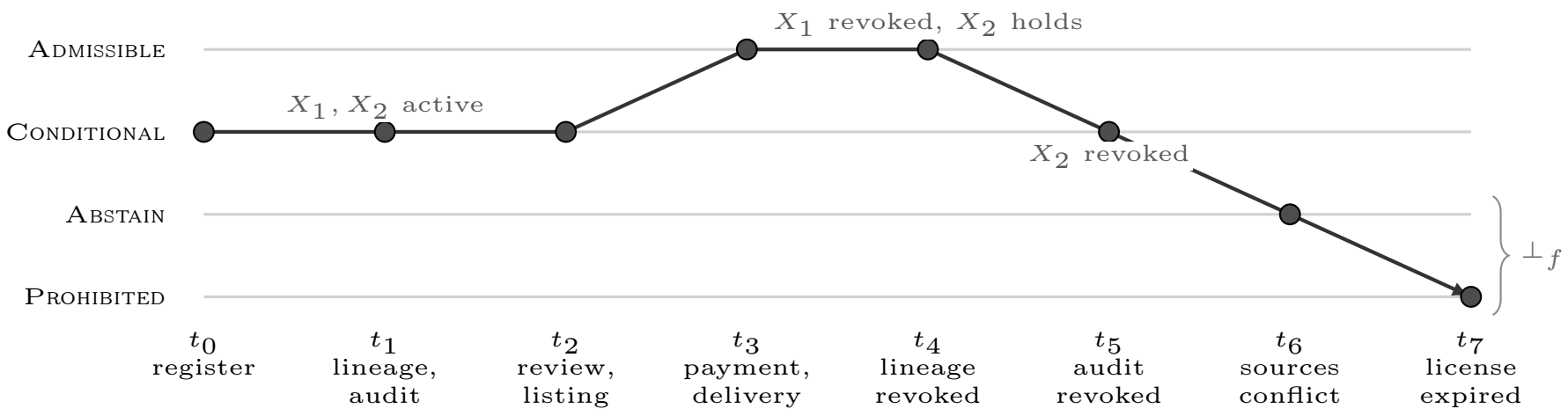

Fig.~\ref{fig:trace} plots the eight-transition trace of a purpose-bound transport-data transaction, with each transition annotated by the event and the resulting claim strength. The walkthrough is constructed for illustration; it is not a deployed trace or a legal validation.

\section{Discussion}

\subsection{Interpretation and engineering implications}

RQ3 concerns executable semantic conformance, not deployed engines or human experts: implementation matches the formal decision table, and the diagnostic negative controls identify omitted guards; they are not competitive baselines. The method sits beside a catalog, policy decision point, API gateway, and audit log without storing raw data: services export claim contracts and evidence references while the assurance layer checks interfaces, returns obligations, and records a trace. Payment completion guarantees neither supply right nor purpose compatibility, so negative results stay actionable --- unresolved rights need validation, conflicts need resolution, expired licenses need re-authorization --- and abstention is explicit, not a failure.

\subsection{Threats to validity}

\textbf{Construct validity.} The small policy and workflow omit legal, privacy, pricing, and derived-data concerns; the trace is constructed, and D1--D4 isolate omitted guards rather than act as baselines. \textbf{Internal validity.} Evaluator and oracle independently implement the same semantics, so agreement checks conformance, not reality; seeded defects show the campaigns discriminate but cannot establish the semantics is right. \textbf{External validity.} The workflow is a requirements abstraction, not a deployed trace; generalization and cost need expert coding and field evaluation. \textbf{Evidence validity.} A passed guard means evidence satisfies a declared contract, not that the source is truthful or the contract exhaustive; common-source dependence needs stronger treatment.

\section{Conclusion}

Evidence-carrying contracts bound a workflow claim by the evidence supporting it, so unresolved and conflicting evidence degrade the claim instead of being silently promoted. The evaluator conforms to the declared semantics across every campaign above, establishing conformance, not operational validity. Future work adds temporal validity, provenance tracing, typed compatibility, and multi-domain evaluation. \textbf{Artifact availability.} The complete reproducibility artifact is publicly archived at \url{https://doi.org/10.5281/zenodo.22737537}.

% Keep the 12-page main-text limit explicit and stable: bibliography begins
% on page 13, while the official LLNCS/CCIS text block remains unchanged.
\clearpage
\bibliographystyle{splncs04}
\bibliography{references}

\end{document}